# Nano-Imaging of Landau-Phonon Polaritons in Dirac Heterostructures

Lukas Wehmeier[1,2*], Suheng Xu[3], Rafael A. Mayer[1], Brian Vermilyea[4], Makoto Tsuneto[1], Michael Dapolito[1,3], Rui Pu[1], Zengyi Du[1], Xinzhong Chen[1,3], Wenjun Zheng[1], Ran Jing[1,5], Zijian Zhou[1], Kenji Watanabe[6], Takashi Taniguchi[7], Adrian Gozar[8,9], Qiang Li[1,5], Alexey B. Kuzmenko[10], G. Lawrence Carr[2], Xu Du[1], Michael M. Fogler[4*], D.N. Basov[3*], Mengkun Liu[1,2*]

**Affiliations:**

[1]Department of Physics and Astronomy, Stony Brook University; Stony Brook, New York 11794, USA.

[2]National Synchrotron Light Source II, Brookhaven National Laboratory; Upton, New York 11973, USA.

[3]Department of Physics, Columbia University; New York, New York 10027, USA.

[4]Department of Physics, University of California at San Diego; La Jolla, California 92093-0319, USA.

[5]Condensed Matter Physics and Materials Science Division, Brookhaven National Laboratory; Upton, New York 11973, USA.

[6]Research Center for Electronic and Optical Materials, National Institute for Materials Science, 1-1 Namiki, Tsukuba 305-0044, Japan.

[7]Research Center for Materials Nanoarchitectonics, National Institute for Materials Science, 1-1 Namiki, Tsukuba 305-0044, Japan.

[8]Department of Physics, Yale University; New Haven, Connecticut 06520, USA.

[9]Energy Sciences Institute, Yale University; West Haven, Connecticut 06516, USA.

[10]Department of Quantum Matter Physics, University of Geneva; 1211 Geneva, Switzerland.

*Corresponding authors: Mengkun Liu: mengkun.liu@stonybrook.edu, D.N. Basov: db3056@columbia.edu, M.M. Fogler: mfogler@ucsd.edu, Lukas Wehmeier: lwehmeier@bnl.gov

Abstract

Polaritons are light-matter quasiparticles that govern the optical response of quantum materials and enable their nanophotonic applications [1–4]. We have studied a new type of polaritons arising in magnetized graphene [5–7] encapsulated in hexagonal boron nitride (hBN) [8–10]. These polaritons stem from hybridization of Dirac magnetoexciton modes of graphene with waveguide phonon modes of hBN crystals. We refer to these quasiparticles as the Landau-phonon polaritons (LPPs). Using infrared magneto nanoscopy, we imaged LPPs and controlled their real-space propagation by varying the magnetic field. These LLPs have large in-plane momenta and are not bound by the conventional optical selection rules, granting us access to the "forbidden" inter-Landau level transitions (ILTs). We observed avoided crossings in the LPP dispersion – a hallmark of the strong coupling regime – occurring when the magnetoexciton and hBN phonon frequencies matched. Our LPP-based nanoscopy also enabled us to resolve two fundamental many-body effects: the graphene Fermi velocity renormalization [11–16] and ILT-dependent magnetoexciton binding energies. These results indicate that magnetic-field-tuned Dirac heterostructures are promising platforms for precise nanoscale control and sensing of light-matter interaction.

**Introduction**

Polaritons are light-matter quasiparticles that play a fundamental role in the optical response of polarizable materials [1–10,17–23]. Phonon-polaritons were studied historically first [24] and they are examples of modes demonstrating strong light-matter coupling. In complex materials polaritons can involve several distinct matter excitations, yielding a rich variety of collective phenomena [25–27,3]. If the optical properties of a material are tunable, polaritons inherit this tunability. For example, the dispersion of plasmon-polaritons in two-dimensional (2D) conductors can be controlled by changing their charge carrier concentration [19–21] or applying an electric current [22,23]. However, attaining strong mode coupling with conducting materials is difficult because of their high electronic losses. Graphene is one of the promising polaritonic platforms because of its low intrinsic electron scattering rate [28] and corresponding high quality factors [8,3,9].

Here, we report the discovery of the Landau-phonon polaritons (LPP) in a 2D graphene-hBN heterostructure. The LPPs result from the hybridization [25–27] of phonon-polaritons of the hBN encapsulating layers [8–10] with Dirac magnetoexcitons [6,7] (or "Landau polaritons" [5]) of charge-neutral graphene [6,7]. LPPs belong under a broader umbrella of magneto-phonon resonance (MPR) effects, resulting from a near coincidence of the energy spacing between a pair of Landau levels and the energy of an optical phonon. We comment on other MPR effects [29–33], such as magneto polarons [29–31], in the outlook. Employing the state-of-the-art magneto scanning near-field optical microscopy (m-SNOM), [6,34–36] we have imaged real-space interference patterns created by the LPPs. We demonstrate that the LPP propagation can be switched on and off using magnetic fields. We have been able to detect as many as six different LPP branches. Several of them originate from optically dark transitions, suggesting that the usual selection rules [12–15,37,38] no longer apply in the extended momentum-frequency space accessible with the m-SNOM. Our high-precision mapping of the LPP dispersion has also enabled us to quantify many-body effects that yield the effective Fermi velocity in graphene. [11–15,39]

Our experimental setup is depicted in Fig. 1a. The experiments involved focusing infrared radiation onto the tip of an atomic force microscope that acted as a scannable nanoscale antenna. Light scattered from the tip carried near-field information to a far-field detector. Another, stationary nanoantenna in the form of a metallic bar deposited on graphene, played the dual role of an electrical contact and a polariton launcher. Both the sample and the m-SNOM resided in an optical cryostat allowing the control of temperature and magnetic field applied in the out-of-plane direction (Supplementary Information and Reference [6]). We present and discuss the results of these measurements below, after we have introduced the necessary theoretical background.

**High-Momentum Magneto-Optics of Graphene**

In a transverse magnetic field, the density of states in graphene splits into Landau levels (LLs) of energy $E_n = \text{sgn}(n)\sqrt{2|n|}(\hbar v_F/l_B)$, where $n = 0, \pm 1, \pm 2, ...$ is the LL index, $v_F$ is the Fermi velocity, $e$ is the elementary charge, $l_B = \sqrt{\hbar/e|B|}$ is the magnetic length, and $B$ is the magnetic field (Fig. 1b). This characteristic square-root $n$- and $B$-dependence is a manifestation of the Dirac-like energy-momentum dispersion of graphene quasiparticles. In a charge-neutral graphene the optical transitions can occur between LLs with indices of opposite sign, $-n \rightarrow n'$, at frequencies $\omega \propto \sqrt{|n|} + \sqrt{|n'|}$. The oscillator strength of each transition is a function of the in-plane momentum $k$. Conventional far-field infrared experiments excite graphene at very small $k$, with a non-negligible oscillator strength only at $|n| - |n'| = \pm 1$. This selection rule at $k \rightarrow 0$ is evident in the nonlocal optical conductivity $\sigma(\omega, k)$ of graphene shown in Fig. 1c (blue curve in Fig. 1d). Such peaks in the optical conductivity Re $\sigma_{xx}$ have been observed in many magneto-optical absorption experiments [12,14,15,37,38,40].

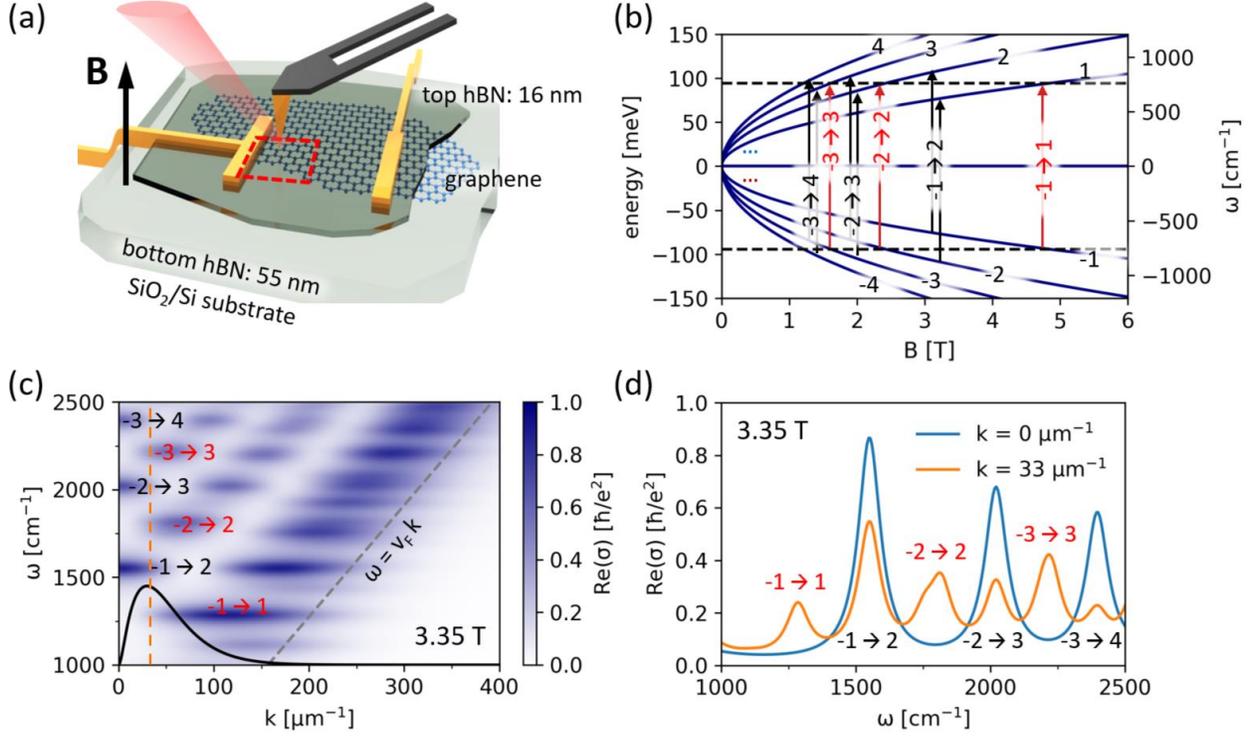

*Figure 1: High-Momentum Magneto-Optics of Graphene. (a) Schematics of our sample and m-SNOM setup. Gold contacts enable transport measurements and gating of graphene and also serve as polariton launchers. (b) Landau level (LL) energy as a function of magnetic field B and LL index $n = 0, \pm 1, \dots, \pm 4$. Black (red) arrows mark $-n \to n \pm 1$ and $-n \to n$ inter-Landau level transitions (ILTs) for a photon energy of $\hbar\omega = 188$ meV ($1519$ cm$^{-1}$). (c) Real part of the graphene conductivity [41] at $B = 3.35$ T as a function of frequency $\omega$ and in-plane momentum $k$ calculated using Fermi velocity $v_F = 1.19 \times 10^6$ m/s and damping $\gamma = 24.3$ cm$^{-1}$. The relevant $-n \to n \pm 1$ ($-n \to n$) ILTs are labeled in black (red). The black bell-shaped curve illustrates the momenta accessible via m-SNOM [20], with $k = 1/r_{\text{tip}} = 33$ $\mu$m$^{-1}$ marked by the vertical dashed line. (d) The line cuts at momenta $k = 0$ and $k = 33$ $\mu$m$^{-1}$ extracted from panel (c).*

The theory also predicts that at finite momenta transitions between any pair of LLs become possible [41]. Among these additional "forbidden" transitions, the first ones to become noticeable as $k$ increases are the $-n \to n$ transitions, see Fig. 1c. Faint signatures of such forbidden modes have been seen in previous far-field experiments [12]. They were attributed to mildly relaxed momentum conservation due to disorder scattering. As discussed below, our experiments have revealed much stronger evidence of the forbidden inter-LL transitions (ILTs), presumably because the requisite large in-plane momenta were created by scattering of light with the tip. Indeed, the forbidden transitions become comparable in strength to the nearby allowed ones at momenta of the order of the inverse magnetic length, e.g., $l_B^{-1} = 71$ μm$^{-1}$ at $B = 3.35$ T. The momentum range important in the m-SNOM is illustrated by the bell-shaped curve in Fig. 1c. For the estimated tip radius of $r_{tip} = 30$ nm, it is centered around $k = 33$ μm$^{-1}$ marked by the vertical dashed line [20]. At such $k$ the forbidden transitions are only slightly weaker than the allowed ones, see the orange line in Fig. 1d. Also, the allowed transitions at nonzero $k$ are diminished with respect to the $k = 0$ case (the blue line in Fig. 1d) to fulfill the optical sum rule.

**Modeling of Polariton Dispersion**

Each ILT gives rise to a collective excitation, which has been previously referred to as a Landau polariton [5] (the term we use here), Dirac magnetoexciton, or magnetoplasmon [6,7]. If the Landau polaritons are tuned in resonance with the hyperbolic phonon-polaritons in hBN by changing the applied magnetic field, the hybrid modes, which are the aforementioned LPPs, can form. We have carried out numerical simulations to model the LPP dispersion expected under our experimental conditions. As customary in near-field studies, we deduce the dispersion of the collective modes from the frequency and momentum-dependent p-polarized reflection coefficient of the sample, $r_p = r_p(k, \omega)$. Figs. 2a-c demonstrate the imaginary part of $r_p$ calculated for three representative values of the magnetic field. The multiple branches of phonon-polaritons in the upper Reststrahlen band of hBN (~1360-1610 cm$^{-1}$) are evident in all three cases [8–10]. Without the magnetic field (Fig. 2a), the charge-neutral graphene influences the response of the heterostructure only weakly, via its "universal" optical conductivity $\sigma = e^2/4\hbar$ [12].

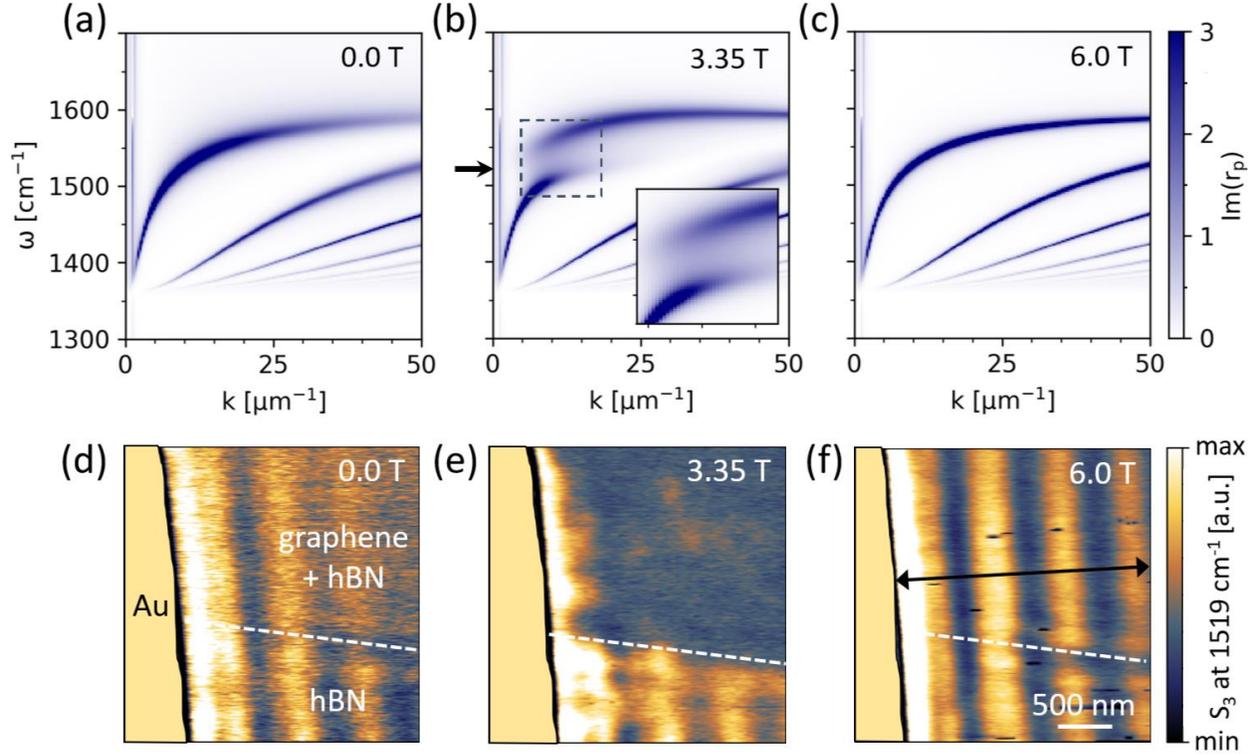

*Figure 2: Hybridization of hBN Phonon Polaritons with Graphene Landau Polaritons, resulting in Landu-Phonon Polaritons (LPPs). (a-c) Calculated LPP dispersion at magnetic fields of 0.0, 3.35, and 6.0 T, respectively. The false color represents $\mathrm{Im}\, r_p(k,\omega)$, the imaginary part of the reflection coefficient for p-polarized light. Graphene is assumed to be charge neutral with a constant LL broadening [12] $\gamma = 24.3$ cm$^{-1}$ and Fermi velocity $v_F = 1.19 \times 10^6$ m/s, the latter being the value extracted from Fig. 4c. Inset in (b) An enlarged view of the region exhibiting strong coupling and an avoided crossing between the Landau and the hBN phonon polaritons; the arrow marks $\omega = 1519\ cm^{-1}$ corresponding to the data in panels d-f. (d-f) Nano-imaging data collected from the region marked by the red rectangle in Fig. 1a at $T = 154$ K and $B = 0.0, 3.35,$ and 6.0 T, respectively. The near-field signal $S_3$ (demodulated at the 3$^{rd}$ harmonic of the tip frequency, refer to Supplementary Information) shows relative differences between regions with and without graphene that strongly depend on the magnetic field. The reduced signal-*

*to-noise ratio in (d), compared to (e-f), is due to a shorter integration time. The double-headed arrow in (f) marks the location of the line scan analyzed further in Fig. 3.*

At 3.35 T (Fig. 2b), the frequency of the $-1 \to 2$ ILT is inside the hBN upper Reststrahlen band, which generates avoided crossings in the polariton dispersion. These features manifest as a coupling and hybridization of the $-1 \to 2$ inter-LL Landau polariton with the hBN phonon polaritons, i.e., the formation of the LPPs. Modeling the system as two coupled harmonic oscillators [42,43] (see Supplementary Information), we extract the mode splitting $\Omega = 43.3$ cm$^{-1}$ at the largest avoided crossing and mode linewidths of $\Gamma_{\text{Landau}} = 50.6$ cm$^{-1}$ and $\Gamma_{\text{hBN}} = 4.0$ cm$^{-1}$ for the uncoupled Landau polariton and hBN phonon polariton, respectively. Hence, the strong coupling criterion $C = \frac{2\Omega^2}{\Gamma_{\text{Landau}}^2 + \Gamma_{\text{hBN}}^2} = 1.5 > 1.0$ is fulfilled at this magnetic field.

At 6.0 T (Fig. 2c), our calculations indicate no ILTs inside the Reststrahlen band, so the phonon-polariton dispersion is again largely unaffected by graphene. Notably, these calculations show that the polariton damping at 6.0 T should be lower than that at 0 T because the LL quantization makes graphene more optically transparent away from the discrete ILT frequencies [12].

**Nano-Imaging of LPP**

We now turn to our experimental nano-imaging results that reveal field-tunable features of LPPs in real space. Figures 2d-f show m-SNOM images acquired at a temperature of 154 K and magnetic fields of $B = 0.0$, 3.35, and 6.0 T, matching Figs. 2a-c, respectively. The incident photon energy is 188 meV (wavenumber ω = 1519 cm$^{-1}$). Note that our scan area contained three different regions: (1) a gold electrode on the left, which served as a polariton launcher; (2) hBN-encapsulated graphene on the top right showing propagating LPP polaritons; (3) hBN without graphene in the bottom right, showing phonon-polariton modes only. At 0.0 T (Fig. 2d) and 6.0 T (Fig. 2f), we observed polariton fringes parallel to the gold electrode in both regions (2) and (3). At 0.0 T, the fringes in the region containing graphene exhibited a higher damping. At 6.0 T, the impact of graphene was minimal. These findings are consistent with our simulations (Fig. 2a,c) and also previous work [12]. On the other hand, at 3.35 T (Fig. 2e), there is a striking contrast between the regions with and without graphene. The polariton propagation in hBN-graphene ceases such that all but the first fringe is suppressed. This gives clear evidence for the existence of the hybridization gap in the LPP dispersion, i.e., the strong mode coupling, predicted by our theoretical calculations (Fig. 2b).

To study the magnetic-field dependence of the LPP dispersion in detail, we obtained a field-tip-position map of the m-SNOM signal (Fig. 3a) by sweeping $B$ from $-6.0$ T to $+6.0$ T. The maps were acquired by performing repeated scans with the tip along lines perpendicular to the gold electrode, as marked by the black arrow in Fig. 2f. At our selected photon energy of 188 meV ($\omega = 1519$ cm$^{-1}$) within the hBN Reststrahlen band, we observe the suppression of the fringes for certain distinct field values, e.g., for the discussed case of $B = 3.35$ T. When approaching such fields from a higher (lower) absolute magnetic field side, the polariton wavelength decreases (increases) along with an overall reduction in near-field signal and a decrease of the propagation length. Figure 3b shows line profiles that have been extracted at $B = 0.0$, $\pm 3.3$, and $\pm 5.8$ T, respectively. While we observe oscillatory polariton fringes at 0.0 and $\pm 5.8$ T, at the $-1 \rightarrow 2$ ILT at $B = \pm 3.3$ T, the fringes are strongly damped, consistent with the predicted observations in Fig. 2. These features are observed for both directions of the magnetic field, $B > 0$ and $B < 0$. We note that the ILTs can also be suppressed by doping graphene off charge neutrality (via the Pauli blocking [12,14,38]), which provides additional opportunities for controlling LPPs (see Supplementary Information).

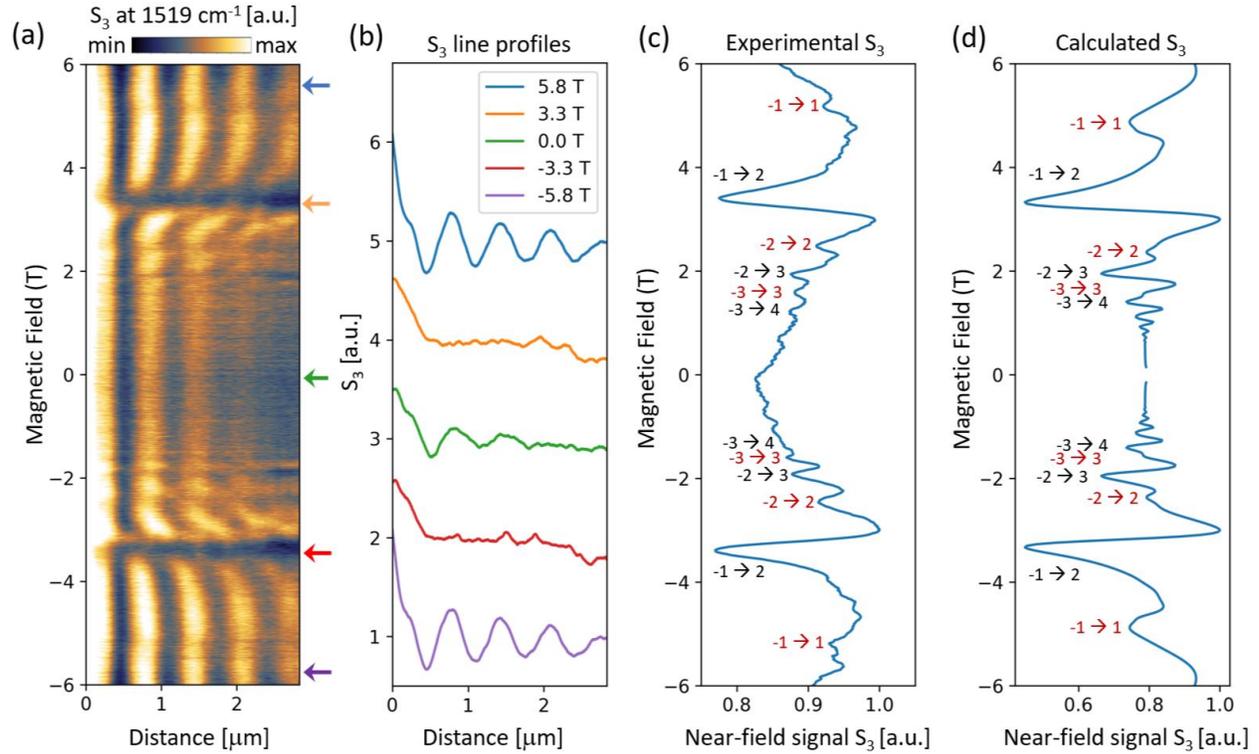

*Figure 3: Magnetic-Field Dependence of the Polariton Dispersion.* *(a)* *Near-field signal $S_3$ acquired via a repeated line scan while sweeping the magnetic field from $-6.0$ to $6.0$ T at a rate of $0.4$ mT/s; measurement were taken at $\omega = 1519$ cm$^{-1}$ and $T = 154$ K. The position of this line scan was perpendicular to the gold contact, which served as a polariton launcher, see Figs. 1a and 2f. (b) Line profiles extracted from (a) at magnetic fields $B = 0.0, \pm 3.3,$ and $\pm 5.8$ T. (c) The near-field signal in (a) averaged over the distance. Minima of the averaged signal are assigned to the ILTs shown by the labels. (d) Calculated near-field signal (Supplementary Information) as a function of magnetic field. Parameter values are chosen to be the same as in Figs. 1c-d and 2a-c.*

Averaging the data in Fig. 3a over the tip position removes the spatial oscillations, which allows us to focus on the $B$-dependence of the signal (Fig. 3c). We can compare the trace in Fig. 3c with the theoretical simulations in Fig. 3d. For simplicity, in these simulations we did not include a separate polariton launcher. Instead, we modeled a more common m-SNOM setup where the tip is the only nanoantenna interacting with the sample, which is uniform and infinite in size. It is fair to compare the predictions of this model with Fig. 3c since in both cases the signal depends only on the magnetic field, not the tip position. We see that Figs. 3c and 3d are in good agreement. The dips in the theoretical curve come from distinct ILTs. Assuming this is also the case in the data, we can label them accordingly. Furthermore, fitting the data to the theory allows us to determine the graphene Fermi velocity, which we discuss in detail below.

We find that the dips corresponding to the $-1 \to 2$ transitions are the strongest in our frequency window, testifying to a strong mode coupling regime (see also Supplementary Information). In addition, we observe clear signatures of several other transitions. They include allowed transitions $-2 \to 3$ and $-3 \to 4$, as well as transitions $-1 \to 1$, $-2 \to 2$, and $-3 \to 3$ forbidden by the standard selection rules [12,15,37,38]. In total, we can resolve six different ILTs in our data. Notably, the forbidden transitions [12,32,44] show up much stronger compared to what was previously seen in far-field infrared spectroscopy [12]. As hypothesized above, this massive breakdown of the selection rules originates from the greater role of high-momentum field components $k \sim l_B^{-1}$ in our m-SNOM measurements (Fig. 1c-d).

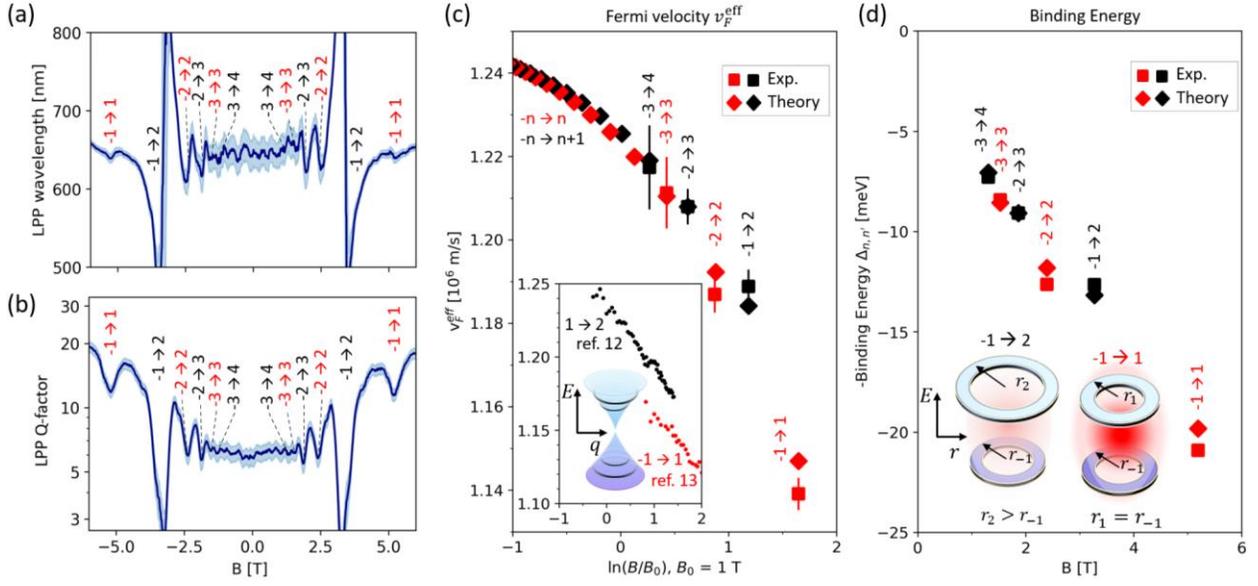

*Figure 4: Magnetic-Field Dependence of LPP Properties and Fermi Velocity Renormalization.*
*(a) Polariton wavelength $\lambda_P$ and (b) polariton quality factor $Q = \operatorname{Re} k / \operatorname{Im} k$ as a function of the magnetic field B. Solid lines show experimental values extracted from Fig. 2; shaded regions show the standard deviation of the measurement. (c) Effective Fermi velocity $v_F^{eff}$ as a function of the logarithmic magnetic field $\ln(B)$, derived for different ILTs (see text): Squares show experimental values derived from (b). Diamonds represent calculated values of $v_F^{eff}$ (see Supplementary Information) [16]. We observe a non-logarithmic trend. Inset: The red (black) points show $v_F^{eff}$ for the -1 → 1 ( -1 → 2 ) ILT measured via Raman spectroscopy [13] (far-field infrared spectroscopy [12]). The tapering shape of the Dirac cone illustrates the Fermi velocity renormalization [11–13], resulting in a logarithmic B-dependence of the far-field data [12,13]. (d) Squares and diamonds show the exciton binding energy $\Delta_{nn'}$ of the Landau polaritons derived from the experiment and theory, respectively. The exciton binding energy is larger for the ILTs with $n' = -n$ compared to those with $n' = -(n \pm 1)$ and generally increases with magnetic field. Inset: The dependence of the exciton binding energy on the magnetic field and type of the ILT can be explained within a semiclassical model where quantized electronic orbitals of the LLs are shaped as narrow rings of radius $r_j = l_B\sqrt{2|j|}$, $j = n$ or $n'$. The magnetoexciton binding energy $\Delta_{nn'}$ (see text) is given by the Coulomb attraction energy of these rings. For a fixed $n$, this binding energy is the largest when the ring radii are equal, at $n' = -n$.*

**Tunability of LPPs**

We have fitted the polaritonic fringes in Fig. 3 to exponentially decaying sine waves $\sim e^{ikx}$, where $k = \operatorname{Re} k + i \operatorname{Im} k$ is the complex polariton momentum (Supplementary Information S2). From this fitting, we deduced the LPP wavelength $\lambda_P = 2\pi/\operatorname{Re} k$ (Fig. 4a) and the quality factor $Q = \operatorname{Re} k/\operatorname{Im} k$ (Fig. 4b). For example, at $B = 0$ T, we found $\lambda_P = 647$ nm and $Q = 12$. As $B$ increases, the crossing of each ILT results in a deep minimum of the $Q$-factor as well as a characteristic change in $\lambda_p$. Within the studied magnetic field range, we have observed a modulation depth of $\lambda_{\max}/\lambda_{\min} \sim 2$ (Fig. 4a) and $Q_{\max}/Q_{\min} \sim 10$ (Fig. 4b). The latter is much larger than the values $Q_{\max}/Q_{\min} \sim 2$ reported for gate-tuning of doped graphene-hBN polaritons [25]. In particular, near the -1 → 2 transition, altering the magnetic field by only 10% changes the $Q$-factor by a factor of five. Therefore, the magnetic field provides a feasible path towards on/off switching of polariton propagation in 2D systems.

**Many-body effects**

Finally, we discuss many-body effects manifested in deviations of the ILT frequencies from the $\sqrt{B}$ - law valid for free Dirac fermions. An alternative description of these deviations is the renormalization of the effective Fermi velocity $v_F^{\text{eff}}$ *defined* by Eq. (1) below. From the minima in the $Q$-factor we can read off the magnetic fields $B$ associated with each ILT and obtain the corresponding field-dependent $v_F^{\text{eff}}$. We find that $v_F^{\text{eff}}$ decreases with $B$ for all transition types following a non-logarithmic B-dependence (squares in Fig. 4c). These values agree surprisingly well with previous far-field infrared [12] and Raman [13] spectroscopy results. Importantly, we extract $v_F^{\text{eff}}$ associated with both allowed and forbidden ILTs with the same measurement. In this regard, the m-SNOM provides a unified approach for LL spectroscopy that lifts many previous limitations.

Our theoretical calculations (Supplementary Information) of the effective Fermi velocity (diamonds in Fig. 4c) show a good agreement with the data utilizing one adjustable parameter, the value of $v_F^{\text{eff}}$ at one specific ILT (here, $-2 \to 3$ gives the best agreement). These calculations also corroborate our experimental observation that the effective Fermi velocity of the $-n \to n$ ILTs is consistently below the trend followed by the $-n \to n \pm 1$ ones.

Our explanation for the above observation of $v_F^{\text{eff}}$ is as follows: Despite its common usage, the term "effective" Fermi velocity is somewhat misleading in the present context. A more accurate statement is that the interaction corrections to the *observed* ILT energy $\hbar\omega$, resulting in $v_F^{\text{eff}}$, include contributions from both the Fermi velocity renormalization (a polaronic effect) and excitonic effects. Namely, $\hbar\omega$ is given by the LL energy difference, $|E_n| + |E_{n'}|$, minus the magnetoexciton binding energy $\Delta_{nn'}$:

$$\hbar\omega \equiv \frac{\hbar v_F^{\text{eff}}}{l_B}\left(\sqrt{2|n|} + \sqrt{2|n'|}\right) = (|E_n| + |E_{n'}|) - \Delta_{nn'}. \tag{1}$$

The LLs $E_n$ in this expression obey the quantization rule $|E_n| = E(q_n)$ where $E = \hbar v_F^{\text{ren}} q$ is the renormalized quasiparticle dispersion and $q_n = l_B^{-1}\sqrt{2|n|}$ is the quantized momentum of a Dirac fermion residing at the *n*th LL (inset Fig. 4c). The effective Fermi velocity $v_F^{\text{eff}}$ is (approximately) equal to the renormalized $v_F^{\text{ren}}$ only if the magnetoexciton binding energy $\Delta_{nn'}$ is neglected. In that case a logarithmic dependence of $v_F^{\text{eff}}$ on $E$ (at fixed $n$ and $n'$) follows from the perturbation theory formula $v_F(E) \approx v_F(\Lambda)\left[1 + \frac{1}{4}\alpha \ln|\Lambda/E| + \cdots\right]$ where $\alpha = e^2/(\kappa\hbar v_F) \ll 1$ is the Coulomb coupling constant, $\Lambda$ is the high-energy cutoff, and $\kappa$ is the effective dielectric constant of the graphene environment [15]. This formula has been derived for graphene in zero magnetic field; however, it remains approximately correct at nonzero $B$ (Supplementary Information), meaning that the renormalized Fermi velocity is first and foremost a function of energy, $v_F^{\text{ren}} = v_F^{\text{ren}}(E)$. Since $E_n$ and $E_{n'}$ are $B$-dependent [12,13], $v_F^{\text{ren}}$ usually acquires a logarithmic $B$-dependence for a given ILT, as found in previous far-field spectroscopy studies [11–13,15,16] (inset in Fig.4c). On the other hand, here we have studied ILTs at a fixed laser frequency so that the transition energy $\hbar\omega \approx |E_n| + |E_{n'}|$ remained the same, being split roughly equally between $|E_n|$ and $|E_{n'}|$. Therefore, in our experiments, the renormalized Fermi velocity $v_F^{\text{ren}} \approx v_F^{\text{ren}}(\hbar\omega/2)$ should have little B-field dependence and the observed variation of $v_F^{\text{eff}}$ (Fig. 4c) should mostly come from the change of magnetoexciton binding energy $\Delta_{nn'}$, which follows non-logarithmic trend with changing magnetic field.

Indeed, our theoretical calculation of the two competing terms, $|E_n| + |E_{n'}|$ and $\Delta_{nn'}$, in Eq. (1) confirms that at fixed $\hbar\omega$, the former gives a nearly constant contribution to $v_F^{\text{eff}}$ for all measured ILTs, so that $v_F^{\text{eff}}$ variation comes from the latter, with characteristic dips occurring at $n' = -n$ points (Supplementary Information). This allows us to extract the binding energy $\Delta_{nn'}$ (Fig. 4d) from $v_F^{\text{eff}}$. The absolute value of $\Delta_{nn'}$ generally increases with magnetic field and is larger for the ILTs with $n' = -n$ compared to those with $n' = -(n \pm 1)$. A simple way to think about the magnetoexciton binding energy $\Delta_{nn'}$ is to imagine that it is equal to the Coulomb attraction energy of two LL orbitals shaped as concentric rings, one with charge $+e$, the other with charge $-e$ (inset Fig. 4d). The ring radii are given by the formula $r_j = |E_j/ev_F^{\text{ren}}B|$, where $j = n$ or $n'$, which is the semiclassical cyclotron radius of a Dirac particle with energy $E_j$. (Note additional relations

$r_j = l_B\sqrt{2|j|} = l_B^2 q_j$.) For a fixed $n$, this attraction energy is the largest when the ring radii are equal, i.e., at $n' = -n$, yielding the lowest $v_F^{\text{eff}}$ at such ILTs.

**Conclusions and Outlook**

Our study has shown that the physics of LPPs is very rich, and it involves simultaneously three types of effects: polaritonic, excitonic, and polaronic. These effects have distinct characteristics: 1) The polaritonic effects change collective mode properties in the heterostructure. 'Forbidden' optical transitions are now accessible in the momentum space offered by m-SNOM. The mode coupling between Landau polaritons (magnetoexcitons) in graphene and phonon polaritons in hBN generates a tunable avoided crossing, which could potentially be further tailored by utilizing other ILTs, e.g., $0 \rightarrow 1$ ILT or multi-layer engineering. 2) The excitonic effects are manifestations of the electron-electron interactions. They lead to a finite binding energy, which also modifies the LPP dispersion. This binding energy can be further tuned via dielectric screening engineering. 3) The polaron effect is another term for the renormalization of the quasiparticle dispersion. Although above we emphasized the role of electron-electron interactions as the reason for the renormalization of the Fermi velocity $v_F^{\text{ren}}$, this interaction is screened by hBN. Hence, the interaction of electrons in graphene with phonons in hBN is included implicitly. In our case $v_F^{\text{ren}}$ does not change much with magnetic field since we keep the incident photon energy the same throughout the experiments. Also, in our calculation of the renormalized Fermi velocity we approximated the hBN dielectric function by its dc ($\omega = 0$) value. Goals for future work can be: i) incorporating more sophisticated theoretical approaches into our model to properly treat electron-phonon coupling and ii) carrying out a frequency-dependent experimental study of LPPs.

As mentioned in the introduction to this paper, LPPs are specific examples of magneto-phonon resonance (MPR) effects. Other known MPR effects include magneto-polarons [29–31], dc magneto-transport oscillations [33], and mode splitting in magneto-Raman spectroscopy [32]. Most of them have been studied in bulk crystals or a single material system. It would be interesting to investigate if these phenomena are affected by finite-momenta LPPs in a 2D heterostructure. Finally, it would be desirable to explore a variety of other nano-magneto-optics phenomena using m-SNOM, including chiral edge magnetoplasmons [45,46], cavity magneto optics [47], the polaritonic Hofstadter butterfly [48], magnetoexcitons of fractional quantum Hall states [49], and collective modes of stripe phases in partially filled Landau levels [50].

**Acknowledgments**

Research on m-SNOM scanning probe platform is supported as part of Programmable Quantum Materials, an Energy Frontier Research Center funded by the U.S. Department of Energy (DOE), Office of Science, Basic Energy Sciences (BES), under award DE-SC0019443. L. W., X.Z.C, M.K.L., and D.N.B. acknowledge U.S. Department of Energy, Office of Science, National Quantum Information Science Research Centers, Co-design Center for Quantum Advantage (C2QA) under contract number DE-SC0012704 for the support of the of the data analysis. M.K.L. acknowledges support from the NSF Faculty Early Career Development Program under Grant No. DMR – 2045425 for the development of the Akiyama probe based SNOM. M.D., M.K.L., and Q.L. acknowledge U.S. Department of Energy, Office of Basic Energy Sciences, Division of Materials Sciences and Engineering, under Contract No. DE-SC0012704 for supporting the sample characterization and theory development. R.A.M. thanks grant 2022/06709-0, São Paulo Research Foundation (FAPESP) for modeling development. K.W. and T.T. acknowledge support from the JSPS KAKENHI (Grant Numbers 20H00354, 21H05233 and 23H02052) and World Premier International Research Center Initiative (WPI), MEXT, Japan. A.B.K. is supported by the Swiss National Science Foundation under Grant No. 200020_201096. X.D. acknowledges support from the NSF under awards DMR-1808491 for sample fabrication. We are grateful to Raul Freitas (Brazilian Synchrotron Light Laboratory, LNLS), Aaron Sternbach and Frank Ruta (both Columbia University) for the helpful discussion, as well as for the helpful discussion and technical support from Jing Li, Heng Wang, and Wenjie Wang.

## Author contributions

L.W., D.N.B, and M.L. conceived the project and designed the experiments. Q.L., G.L.C., X.D., M.M.F., D.N.B., and M.L. supervised the project. R.P., K.W., T.T., and X.D. prepared the devices. L.W., M.T., Z.D., and M.L. performed the experimental measurements with support from M.D., W.Z., and Z.Z. L.W. and M.L. analyzed the experimental data with input from S.X., R.A.M., R.J., A.B.K., G.L.C., M.F., and D.N.B. B.V. and M.F. developed the theoretical description. S.X. and R.A.M. developed the numerical simulation with input from L.W., X.C., R.J., M.F., and M.L. L.W., M.M.F., and M.L. co-wrote the manuscript with input from all co-authors.

## Competing interests

M.D., X.C., and M.L. have a patent pending regarding magneto scanning near-field optical microscopy. All other authors declare no competing interests.